\documentclass[preprint]{revtex4}
\usepackage{epsfig}

\begin{document}

\title{Resonant x-ray scattering study on multiferroic
  $\rm\bf BiMnO_3$}

\author{C.-H. Yang, J. Koo, C. Song, T. Y. Koo*, K.-B. Lee, and Y. H. Jeong}
\affiliation{\it Department of Physics \& Electron Spin Science
Center and *Pohang Accelerator Laboratory, Pohang University of
Science and Technology, Pohang, Korea}

\date{\today}

\begin{abstract}
Resonant x-ray scattering is performed near the Mn $K$-absorption
edge for an epitaxial thin film of  $\rm BiMnO_3$. The azimuthal
angle dependence of the resonant (003) peak (in monoclinic
indices)  is measured with different photon polarizations; for the
$\sigma\rightarrow\pi'$ channel a 3-fold symmetric oscillation is
observed in the intensity variation, while the
$\sigma\rightarrow\sigma'$ scattering intensity remains constant.
These features are accounted for in terms of the peculiar ordering
of the manganese 3$d$ orbitals in BiMnO$_3$. It is demonstrated
that the resonant peak persists up to 770 K with an anomaly around
440 K; these high and low temperatures coincide with the
structural transition temperatures, seen in bulk, with and without
a symmetry change, respectively. A possible relationship of the
orbital order with the ferroelectricity of the system is
discussed.
\end{abstract}


\maketitle

\vskip2pc

\narrowtext

\section{Introduction}

Recently multiferroics with the coexistence of ferroelectricity
and magnetism have been intensively studied not only due to
potential  applications but also from basic
interest~\cite{Fiebig}. In particular, BiMnO$_3$ is regarded as a
true multiferroic material with both ferromagnetism and
ferroelectricity. $\rm BiMnO_3$ possesses a distorted perovskite
structure with monoclinic symmetry due to highly polarizable
bismuth ions and Jahn-Teller (JT) active manganese ions. Its
magnetic moment certainly arises from the $3d^4$ electrons of $\rm
Mn^{3+}$ ion, while the ferroelectric polarization does from the
Bi $6s^2$ lone pair~\cite{NAHill}. The saturated magnetization for
a polycrystalline sample was reported to be 3.6 $\mu_{\rm B}/{\rm
Mn}$ and the ferromagnetic transition temperature 105
K~\cite{Chiba}. While the ferroelectric remanent polarization of
BiMnO$_3$ was measured to be 0.043 $\rm \mu C/cm^2$ at 200 K for a
polycrystalline sample~\cite{Moreira450K}, there seems to be no
consensus on the ferroelectric transition temperature $T_{\rm
FE}$. Moreira dos Santos $et$ $al.$, for example, reported $T_{\rm
FE}\,\sim\,$450 K where a reversible structural transition occurs
without a symmetry change~\cite{Moreira450K}; however, Kimura $et$
$al.$ suggested $T_{\rm FE}\,\sim\,$770 K where a
centrosymmetric-noncentrosymmetric structural transition occurs
~\cite{TKimura}.

In BiMnO$_3$ orbital degrees of freedom are also active besides
magnetic and electric ones; the half-filled $e_g$ level of $\rm
Mn^{3+}$ ion induces the JT distortion and an anisotropic charge
distribution with the $3d_{\rm 3z^2-r^2}$ orbital prevails
locally. Then long range orbital order, which is compatible with
the structure, would exist in the system. The detailed structure
of  BiMnO$_3$ at room temperature was investigated by neutron
powder diffraction and the compatible orbital order was
proposed~\cite{AMoreira}. It is noted that no noticeable
structural change associated with the ferromagnetic transition was
observed. Regarding the orbital order in $\rm BiMnO_3$, there
remain unresolved issues such as  the temperature dependence of
the degree of ordering at high temperatures and the determination
of the transition temperature itself, $T_{\rm o}$. Moreover, the
relationship between the ferroelectricity and the orbital order in
this multiferroic compound is clearly of interest.

In order to address these issues we took advantage of resonant
x-ray scattering (RXS) which can probe local environments of a
specific element. RXS occurs as a second order process by first
exciting a core electron into an empty valence state, as the
incident photon energy is tuned to an absorption edge, and then
de-exciting the electron back to the original state. Direct
involvement of the valence state in the scattering process usually
gives rise to enhanced sensitivity to anisotropic charge
distributions, and the atomic scattering factor is no longer a
scalar but takes a tensorial form called  anisotropic tensor
susceptibility (ATS). ATS then allows reflections at some of
normally forbidden positions. Pioneering works regarding ATS were
done in the
eighties~\cite{Templeton,Dmitrienko,Blume,HoDGibbs,Finkelstein},
and more recently Murakami $et$ $al.$ successfully applied RXS to
several orbital ordered compounds~\cite{LSMOMurakami,YMurakami}.
Here we wish to present detailed RXS measurements  for $\rm
BiMnO_3$ to reveal a role of the orbital order in this important
multiferroic compound. The energy profile, azimuthal angle
dependence, and temperature dependence of an ATS reflection peak
will be presented and discussed.

\section{Experimentals}

In order to carry out RXS measurements single crystalline samples
are needed, but BiMnO$_3$ is notoriously difficult to synthesize
in a single crystalline form. Thus we resorted to epitaxial films;
an epitaxial film of BiMnO$_3$ was grown on a SrTiO$_3$(111)
substrate (cubic Miller indices) by the pulsed laser deposition
technique. The thickness of the film was approximately 40 nm. The
detailed growth conditions and structural characterizations are
presented elsewhere~\cite{Yang}. The crystal structure of the thin
film was still monoclinic as in bulk. The growth direction or
surface-normal direction, which is parallel to the [111] direction
of the substrate, is perpendicular to the monoclinic $ab$-plane.
Note that with this orientation the modulation vector of the
orbital ordering of BiMnO$_3$ is along the surface-normal. (See
below.) Thus the orientation of the film was deliberately chosen
in order to make the axis of azimuthal rotation (direction of the
scattering vector) coincide with the surface-normal, so that
azimuthal scans are easily performed. Resonant x-ray scattering
experiments were performed on a 4-circle diffractometer installed
at the 3C2 beam line in the Pohang Light Source. The photon energy
near Mn $K$-edge was calibrated using a manganese foil and the
resolution was about 2 eV. The $\sigma$ polarized photon (electric
field perpendicular to the scattering plane) was injected after
being focused by a bent mirror and monochromatized by a Si(111)
double crystal. The linearly polarized $\sigma'$ (electric field
perpendicular to the scattering plane) and $\pi'$ (electric field
in the scattering plane) components of scattered photons were
separated
with a flat pyrolithic graphite crystal with the scattering angle
of 69 $^{\rm o}$ for the Mn $K$-edge.

\section{Crystal Structure and Structure factor of ${\bf\rm BiMnO_3}$}

\begin{figure}
\vbox{ \centerline{\epsfig{figure=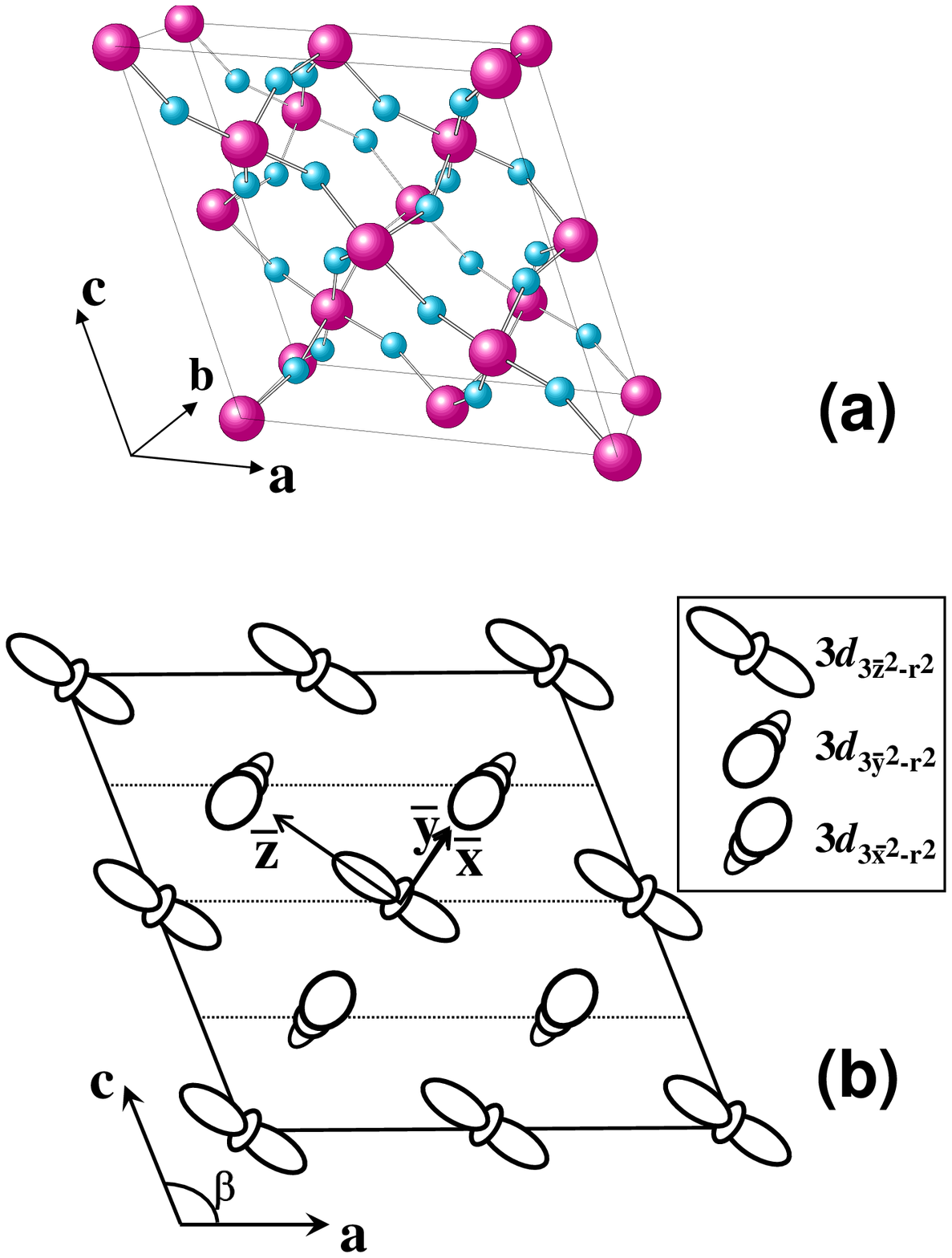, width= 8cm, angle=0}}
}

\caption{(a) The monoclinic cell of BiMnO$_3$. Larger spheres
represent manganese ions and smaller ones oxygen ions. Bismuth
ions are omitted for simplicity. The crystallographic axes for the
monoclinic unit cell are denoted with $a$, $b$, and $c$. The Mn
$ab$-planes are located at $z=0$, $z=\frac{1}{4}-\epsilon$,
$z=\frac{1}{2}$, and $z=\frac{3}{4}+\epsilon$ along the $c$-axis.
$\epsilon$ denotes a small displacement. The angle ($\beta$)
between the $a$- and $c$-axis is approximately 110$^{\rm o}$. (b)
The orbital ordering in the unit cell. The axes for the
pseudo-cubic cell are represented by $\bar{x}$, $\bar{y}$, and
$\bar{z}$. The $ab$-planes at $z=0$ and $z=\frac{1}{2}$ contain
the manganese ions with the orbital ($d_{3\bar{z}^2-r^2}$)
elongated to the pseudo-cubic $\bar{z}$-axis, while the
$ab$-planes at $z=\frac{3}{4}+\epsilon$ and
$z=\frac{1}{4}-\epsilon$ do only $d_{3\bar{y}^2-r^2}$ and
$d_{3\bar{x}^2-r^2}$ orbitals, respectively. }

\label{BMOfig}
\end{figure}

From the neutron powder diffraction with bulk samples at room
temperature the space group (monoclinic $C2$) and local bond
lengths and angles of BiMnO$_3$ were all
determined~\cite{AMoreira}, and it turned out that this monoclinic
structure is maintained in thin films~\cite{Yang}. The
crystallographic unit cell is illustrated in Fig.~\ref{BMOfig}(a).
Note that $a$, $b$ and $c$ denote the crystallographic axes (and
the lattice parameters) for the monoclinic unit cell, and
$\bar{x}$, $\bar{y}$ and $\bar{z}$ represent the axes of the
perovskite pseudo-cubic cell. The structure also determines  the
orientation of JT-distorted MnO$_6$'s and the compatible ordered
pattern of the $e_g$ orbitals of Mn$^{3+}$. The monoclinic
$ab$-planes at $z=0$ and $z=\frac{1}{2}$ contain manganese ions
with the $e_g$ orbital elongated to the pseudo-cubic
$\bar{z}$-axis $d_{3\bar{z}^2-r^2}$, while the orbitals in the
$ab$-planes at $z=\frac{3}{4}+\epsilon$ and
$z=\frac{1}{4}-\epsilon$ are of type $d_{3\bar{y}^2-r^2}$ and
$d_{3\bar{x}^2-r^2}$, respectively. Here $z$ denotes the
coordinate along the $c$-axis and $\epsilon$ designates a small
displacement. As illustrated in Fig.~\ref{BMOfig}(b), this orbital
order is rather peculiar in the sense that each plane parallel to
the $ab$-plane and separated by $c/4$ contains only one kind of
the $e_g$ orbital within the plane and the stacking sequence goes
as
$d_{3\bar{z}^2-r^2}$/$d_{3\bar{x}^2-r^2}$/$d_{3\bar{z}^2-r^2}$/$d_{3\bar{y}^2-r^2}$/$\ldots$($\equiv
\rm Z/X/Z/Y/\ldots$). This long range orbital order corresponds to
the superstructure of the system, and the periodic arrangement of
the anisotropic charge distributions would give rise to measurable
intensities at (00L) ATS reflections with odd L where normal
charge scattering is almost absent (not completely absent because
of nonzero $\epsilon$).

With the crystal structure just described the structure factor
$\mathcal{S}_{\rm (00L)}$ for BiMnO$_3$ (odd L) can be written as
\begin{equation} \label{eqSF}
  \mathcal{S}_{\rm (00L)} \,=\, 2 \mathcal{F}_z e^0 + 2 \mathcal{F}_x e^{i 2\pi {\rm L}(\frac{1}{4}-\epsilon)} + 2 \mathcal{F}_z e^{i 2\pi {\rm
 L}
\frac{1}{2}} + 2 \mathcal{F}_y e^{i 2\pi {\rm
L}(\frac{3}{4}+\epsilon)} + S({\rm Bi, O})
\end{equation}
where $\mathcal{F}_x$, $\mathcal{F}_y$, and $\mathcal{F}_z$
represent the ATS of the manganese ion with anisotropic $e_g$
orbitals elongated along $\bar{x}$, $\bar{y}$, and $\bar{z}$,
respectively, and $S({\rm Bi, O})$ is symbolic of the contribution
of Thomson scattering (normal charge scattering) due to bismuth
and oxygen ions. $\mathcal{F}_z$, for example, may be written,
based on the pseudocubic axes shown in Fig.~\ref{BMOfig}, as
follows:
\begin{equation}
\mathcal{F}_z = \left(
\begin{array}{ccc}
  f_{\bot} & 0 & 0 \\
  0 & f_{\bot} & 0 \\
  0 & 0 & f_{\parallel} \\
\end{array}
\right)\end{equation} where $f_{\bot}$ and $f_{\parallel}$ stand
for the resonant scattering amplitude with photon polarization
perpendicular and parallel to the $e_g$ elongation direction,
respectively.

Then Eq.~(\ref{eqSF}) becomes

\begin{equation}\label{eqSF2}
  \mathcal{S}_{\rm (00L)} \,\simeq\, \underbrace{2(-1)^{\frac{\rm L-1}{2}}i f_{\Delta} \left(
\begin{array}{ccc}
  1 & 0 & 0 \\
  0 & -1 & 0 \\
  0 & 0 & 0 \\
\end{array}
\right)}_{\rm ATS ~~contribution} + \underbrace{8\pi {\rm
L}(-1)^{\frac{\rm L-1}{2}}\epsilon\, f_{\rm Mn} + S({\rm Bi,
O})}_{\rm Thomson ~~scattering}\,,
\end{equation}
where $f_{\rm Mn}$ denotes the isotropic part of the scattering
factor of manganese, and $f_{\Delta}$ $(\equiv\,
f_{\parallel}-f_{\bot})$ denotes the anisotropic portion.

Note that these (00L) reflections with odd L are allowed even in
the absence of resonant scattering because the slight displacement
$\epsilon$ of Mn ions from the exact position prohibits perfect
cancellation of the various contributions. $S({\rm Bi, O})$ in
Eq.~(\ref{eqSF}) probably also contributes. ATS reflections are
usually so small that the observation of RXS is normally difficult
in the presence of the background. The anomalous ATS reflections
from the BiMnO$_3$ film, however, were measurable because the
charge scattering background at the (00L) peaks turned out to be
reduced by a factor of $10^{-5}$ compared to the intensity of the
fundamental (004) peak.

\section{Resonant x-ray scattering}
\begin{figure}
\vbox{ \centerline{\epsfig{figure=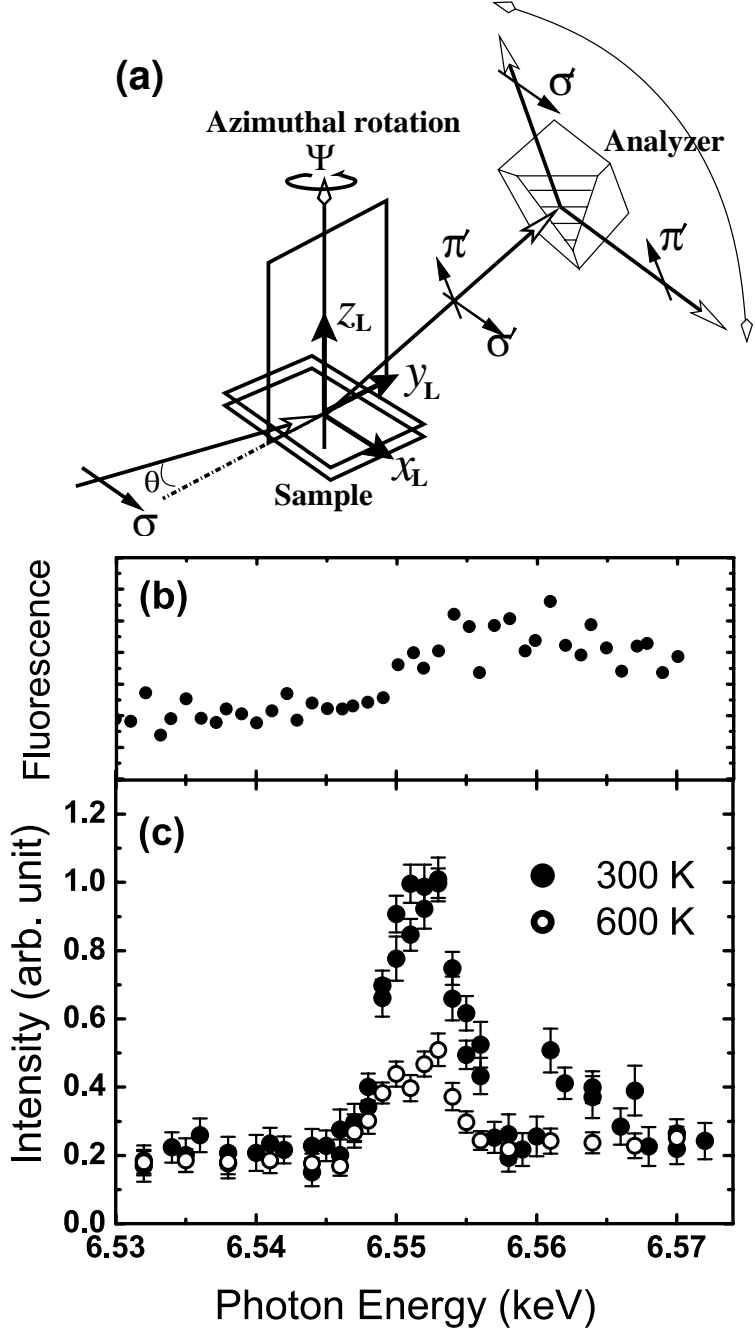, width= 10cm, angle=0}}
}

\caption{(a) The experimental configuration for resonant x-ray
scattering. $\sigma$, $\sigma'$, and $\pi'$  represent the
polarization of the incident and scattered photons. A flat
pyrolithic graphite crystal is used as an analyzer. The cartesian
coordinate ($x_{\rm L}y_{\rm L}z_{\rm L}$) represents the
laboratory frame. (b) The fluorescence data represent the Mn $K$
absorption edge. These are collected at 300 K. (c) The energy
profiles for the (003) peak measured without an analyzer crystal.
The integrated intensities were not corrected for absorption and
were obtained at $\psi=0$.}

\label{BMOENE}
\end{figure}

In Fig. 2(a) schematically shown is the experimental configuration
for RXS. Again $\sigma$, $\sigma'$, and $\pi'$  represent the
polarization of the incident and scattered photons. Azimuthal
rotation and analyzer configuration are also indicated in the
figure. The fluorescence was measured as a function of photon
energy to determine the Mn $K$ absorption edge. The absorption
from the manganese ion results in the uprise of fluorescence above
6.55 keV indicating the Mn $K$ absorption edge as presented in
Fig.~\ref{BMOENE}(b). The integrated intensity of the (003)
reflection was measured as a function of photon energy as shown in
Fig.~\ref{BMOENE}(c). As expected from the resonant nature, the
intensity is sharply enhanced at 6.552 keV in contrast to the
normal charge scattering intensity which would decrease instead
from absorption. Note that at off-resonance energies, appreciable
integrated intensity remains, instead of decaying to zero, as a
result of normal charge scattering described above. Another
interesting point is that, as shown in the two energy profiles at
different temperatures (300 K, 600 K), the resonant contribution
decreases with temperature in contrast to the off-resonant part
which remains nearly constant.

\begin{figure}
\vbox{ \centerline{\epsfig{figure=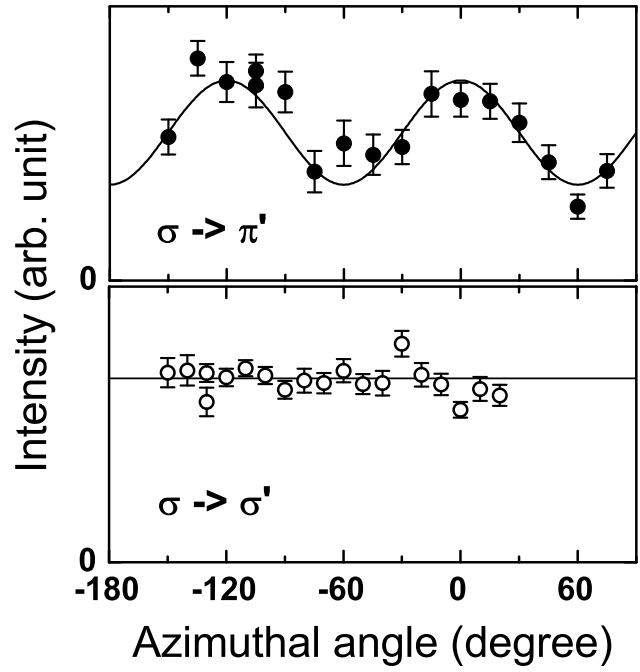, width= 10cm, angle=0}}
}

\caption{The azimuthal angle dependence of the $(003)$ integrated
intensity for the two channels: $\sigma\rightarrow\pi'$ and
$\sigma\rightarrow\sigma'$. The lines in the figure were obtained
from the calculation of resonant x-ray scattering based on the
orbital ordering. See text.}

\label{BMOPol}
\end{figure}

Even if the peak occurred at the expected position in the
reciprocal space, this fact alone may not be sufficient to
conclude that orbital ordering is indeed the origin of the
scattering. The hall mark of the resonant scattering due to
orbital ordering would be the polarization dependence and
azimuthal variation of the peak intensity. In contrast to
conventional x-ray scattering which neither depends on the
azimuthal angle nor shows polarization conversion, resonant
scattering would exhibit the characteristic oscillations,
depending on the photon polarization, with respect to the
azimuthal angle, i.e., the rotation angle around the scattering
vector. Thus in order to further confirm the orbital order as the
origin of this resonant reflection, the azimuthal angle dependence
and polarization analysis were carried out.  In Fig.~\ref{BMOPol}
presented is the azimuthal angle dependence of the (003)
integrated intensity normalized with the fundamental (004) peak
intensity in order to remove sample-shape effects. The zero of the
azimuthal angle $\psi$ is defined when ${\bf k_i}\times{\bf k_f}$
coincides with the $+b$ axis. ${\bf k_i}$ and ${\bf k_f}$, of
course, represent the propagation vector of incident and scattered
beams, respectively. For the $\sigma\rightarrow\pi'$ channel, a
clear three-fold oscillation is observed, while the intensity does
not exhibit any azimuthal dependence for
$\sigma\rightarrow\sigma'$. No doubt this polarization dependence
and the three-fold oscillation seen in the $\sigma\rightarrow\pi'$
channel are due to orbital ordering. It is noted, however, that
these results are not trivially connected to the orbital order of
BiMnO$_3$. What is required is one further step of taking into
consideration the effects of twins which always exist in the film;
the detailed accounts are given in a separate section which
follows.

\begin{figure}
\vbox{ \centerline{\epsfig{figure=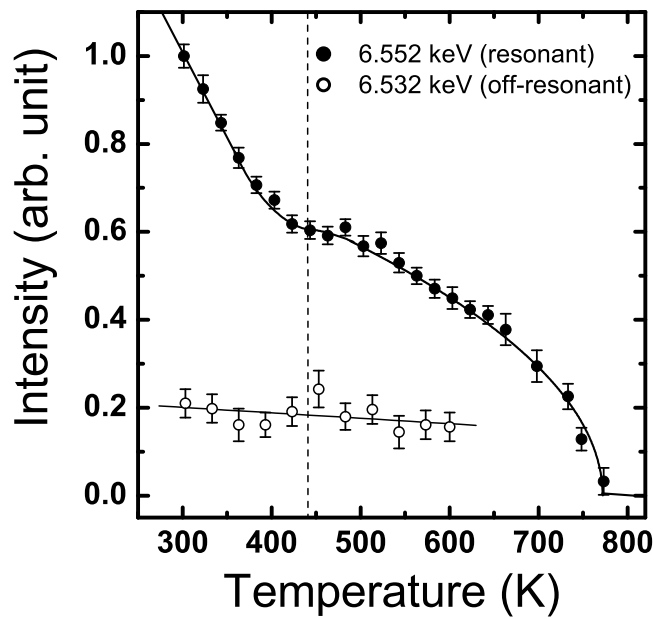, width= 10cm,
angle=0}} }

\caption{The (003) integrated intensities are plotted as a
function of temperature at two different photon energies. The
intensities were collected at $\psi=0$. The intensity measured at
photon energy 6.552 keV are normalized with the fundamental (004)
reflection. Solid lines are a guide for the eye. The intensity
disappears around 770 K. The dashed line indicates 440 K, where an
abrupt increase in the measured intensity seems to start when
temperature is reduced. }

\label{BMOTemp}
\end{figure}

The temperature variation of the orbital order was elucidated by
measuring the (003) integrated intensities at two different photon
energies, 6.532 keV and 6.552 keV, as shown in Fig.~\ref{BMOTemp}.
While the data obtained at off-resonance energy 6.532 keV are
solely due to Thomson scattering (normal charge scattering), those
RXS data at 6.552 keV are caused by the ATS reflection. The
temperature profile of the RXS data shows two distinct features:
one is that the peak intensity decreases rather rapidly with
temperature and disappears  around 770 K. The other is that an
anomaly in the temperature variation appears around 440 K, below
which there is a sudden upturn in the slope. The former
corresponds to the transition temperature of the
centrosymmetric-noncentrosymmetric structural phase transition in
bulk~\cite{TKimura}. The (003) integrated intensity at
off-resonance energy 6.532 keV shows a normal behavior, i.e.,
decreases with temperature due to thermal vibrations, and will
disappear at the structural phase transition point. The lower
temperature 440 K also coincide with the transition temperature
found in bulk; in this case the phase transition involved is the
one without a symmetry change. These two temperatures are
currently under controversy as a  point below which
ferroelectricity first appears. At high temperatures, it should be
noted, it is not trivial at all to prove ferroelectricity of a
material. At any rate, the present RXS results prove that there is
a strong connection between the ferroelectricity and orbital order
in BiMnO$_3$. Of course, the orbital order also determines the
exchange interactions and consequently the magnetic ordering.

\section{Calculation of azimuthal angle dependence}

The azimuthal angle dependence of the scattering intensity from
the anomalous part proportional to $2(-1)^{\frac{L-1}{2}}i$ of
Eq.~(\ref{eqSF2}) can be calculated for the experimental
configuration shown in Fig.~2(a). The conversion matrix
($\mathcal{V}$) from the laboratory frame ($x_{\rm L}y_{\rm
L}z_{\rm L}$) to the pseudo-cubic crystallographic frame
($\bar{x}\bar{y}\bar{z}$) is given by
\begin{equation}
\mathcal{V} = \frac{1}{\sqrt{6}}\left(
\begin{array}{ccc}
 -\sqrt{3} & -1 & \sqrt{2} \\
 \sqrt{3} & -1 & \sqrt{2} \\
 0 & 2 & \sqrt{2} \\
 \end{array}
 \right).
 \end{equation}
The polarization vector of the $\sigma$, $\sigma'$, and $\pi{'}$
polarizations under azimuthal rotation by the amount of $\psi$ can
be written in the laboratory frame as follows:
\begin{eqnarray}
\left|\sigma\right>_{\psi} &=& \left|\sigma'\right>_{\psi} = \mathcal{R}(\psi)\left(\begin{array}{c} 1 \\
0
\\ 0
\\\end{array}\right), \\
\left|\pi'\right>_{\psi} &=& \mathcal{R}(\psi)\left(\begin{array}{c} 0 \\ -\sin{\theta} \\
\cos{\theta} \\\end{array}\right),
\end{eqnarray}

where $\theta$ ($\equiv {2\theta}/{2}$) is the Bragg angle, and
$\mathcal{R}(\psi)$ is the azimuthal rotation matrix given by
\begin{equation}
\mathcal{R}(\psi) = \left(
\begin{array}{ccc}
  \cos{\psi} & -\sin{\psi} & 0 \\
  \sin{\psi} & \cos{\psi} & 0 \\
  0 & 0 & 1 \\
\end{array}
\right).
\end{equation}

From these quantities the scattering intensity caused by ATS with
$\sigma$ as the polarization of the incident photons is easily
calculated for the two channels, $\sigma\,\rightarrow\,\sigma'$
and $\sigma\,\rightarrow\,\pi'$:
\begin{equation}
 I_{\sigma\rightarrow\sigma', \pi'} = \left|\left<\sigma', \pi'\right|_{\psi}\mathcal{V}^{\dag} \left(
 \begin{array}{ccc}
   2 f_{\Delta} & 0 & 0 \\
   0 & -2 f_{\Delta} & 0 \\
   0 & 0 & 0 \\
 \end{array} \right) \mathcal{V} \left|\sigma\right>_\psi \right|^2
\end{equation}
Simplifying this equation yields the scattering intensities as
\begin{eqnarray}
I_{\sigma\rightarrow\sigma'} &=&
\frac{4}{3}f_{\Delta}^2\sin^2{2\psi}, \\
I_{\sigma\rightarrow\pi'} &=&
\frac{4}{3}f_{\Delta}^2(\sqrt{2}\cos{\theta}\cos{\psi} +
\cos{2\psi}\sin{\theta})^2.
\end{eqnarray}

There is one more factor to be considered before comparison with
the experimental data is attempted, that is, the existence of
twins has to be taken into account. The films grown on
SrTiO$_3$(111) would include three kinds of stacking sequences of
orbitals , i.e., $\rm Z/X/Z/Y/\ldots$, $\rm X/Y/X/Z/\ldots$, and
$\rm Y/Z/Y/X/\ldots$. Note that the growth direction, which  is
the cubic [111] direction, is perpendicular to monoclinic (001) or
$ab$- planes. Thus the measured intensities would be the ones
obtained after averaging for the twins:
\begin{eqnarray}
I_{\sigma\rightarrow\sigma'}^{avr} &=& \frac{2}{3}f_{\Delta}^2, \\
I_{\sigma\rightarrow\pi'}^{avr} &=&
\frac{f_{\Delta}^2}{3}(3+\cos{2\theta} +
2\sqrt{2}\sin{2\theta}\cos{3\psi}).
\end{eqnarray}
After twin averaging, the calculated expressions as a function of
$\psi$ agree with the experimental results depicted in the
Fig.~\ref{BMOPol}. The data for the $\sigma\rightarrow\sigma'$
channel do not show any $\psi$ dependence, whereas for the
$\sigma\rightarrow\pi'$ channel a characteristic 3-fold symmetry
is exhibited.

\section{Conclusions}

The resonant x-ray scattering technique was used with epitaxial
thin films  to probe the orbital ordering of a multiferroic
compound, $\rm BiMnO_3$. ATS reflection was observed at (003) in
the reciprocal space as expected, and the characteristic
three-fold oscillation showed up as a function of azimuthal angle.
The experimental results agree with the calculation based on the
orbital ordering when twin effects are taken into account. The
peak intensity was followed as a function of temperature to high
temperatures; it rapidly decreases with temperature and disappears
at $T_{\rm o}\,\approx\,$770 K. Thus orbital ordering occurs
concurrent with the centrosymmetric-noncentrosymmetric structural
phase transition seen in bulk.  The peak intensity also shows an
anomaly at 440 K, where again a phase transition was observed in
bulk. Although there is no consensus about the ferroelectric phase
transition point in BiMnO$_3$, being either 440 K or 770 K, the
orbital ordering is closely connected to the ferroelectricity in
the system in either case.

\acknowledgments
 We wish to thank the financial supports from
SRC-MOST/KOSEF. Synchrotron measurements were done at the Pohang
Light Source operated by POSTECH and MOST.

\newpage

\end{document}